\documentclass[aps,prc,superscriptaddress,showpacs,floatfix,nofootinbib,twocolumn]{revtex4-1}

\usepackage{amsmath,graphicx,float,hyperref}

\newcommand{\trento}{T$\mathrel{\protect\raisebox{-2.1pt}{R}}$ENTo}

\begin{document}
\title{Hydrodynamic predictions for 5.44 TeV Xe+Xe collisions}

\author{Giuliano Giacalone}
\affiliation{Institut de physique th\'eorique, Universit\'e Paris Saclay, CNRS, CEA, F-91191 Gif-sur-Yvette, France}
\author{Jacquelyn Noronha-Hostler}
\affiliation{Department of Physics and Astronomy, Rutgers University,
  Piscataway, NJ 08854, USA}
\author{Matthew Luzum}
\affiliation{Instituto de F\'{i}sica, Universidade de S\~{a}o Paulo, C.P.
66318, 05315-970 S\~{a}o Paulo, SP, Brazil}
\author{Jean-Yves Ollitrault}
\affiliation{Institut de physique th\'eorique, Universit\'e Paris Saclay, CNRS, CEA, F-91191 Gif-sur-Yvette, France} 
%\date{\today}

\begin{abstract}
We argue that relativistic hydrodynamics is able to make robust
predictions for soft particle production in Xe+Xe collisions at the 
CERN Large Hadron Collider (LHC). 
The change of system size from Pb+Pb to Xe+Xe provides a unique 
opportunity to test the scaling laws inherent to fluid
dynamics. 
Using event-by-event hydrodynamic simulations, we make quantitative predictions for several observables:
mean transverse momentum, anisotropic flow coefficients, and their fluctuations.
Results are shown as function of collision centrality.
\end{abstract}

\maketitle
\section{Introduction}
Relativistic hydrodynamics has proven successful in describing the evolution of the system formed in ultrarelativistic nucleus-nucleus collisions~\cite{Heinz:2013th,Bernhard:2016tnd,McDonald:2016vlt,Niemi:2015qia,Niemi:2015voa}.
Typical signatures of hydrodynamics, such as elliptic flow~\cite{Ollitrault:1992bk} or jet quenching~\cite{Wang:2004dn}, depend on simple macroscopic properties of the quark-gluon medium, such as its shape and size. %\cite{Bozek:2013uha,Nagle:2013lja,Goldschmidt:2015kpa,Wang:2004dn}. 
%JYO: these references are about small systems and irrelevant in this context
Challenging tests of this macroscopic description can therefore be performed by studying how observables evolve under variations of the medium geometry.
To achieve this, one can either study a given colliding system in various centrality windows, or collide different species of nuclei with significant variation in the mass numbers.

So far, in addition to $p$+$p$ collisions, $^{208}{\rm Pb}$+$ ^{208}{\rm Pb}$~\cite{Muller:2012zq} and $p+ ^{208}$Pb collisions have been carried out at the CERN Large Hadron Collider (LHC). 
While there seems to be a consensus that hydrodynamics applies to collisions between heavy nuclei, there is still a debate as to whether typical signatures of a quark-gluon plasma observed in small systems~\cite{Bozek:2012gr,Kozlov:2014fqa,Weller:2017tsr} (e.g. the ridge~\cite{Khachatryan:2010gv,CMS:2012qk,Abelev:2012ola}, or multiparticle correlations~\cite{Aad:2013fja,Khachatryan:2015waa,Khachatryan:2016txc}) are also of hydrodynamic origin, or not~\cite{Dusling:2013qoz,Iancu:2017fzn,Dusling:2017aot,Esposito:2015yva,Blok:2017pui}.
On October 9th 2017, an eight-hour run of collisions between $^{129}$Xe nuclei at a center-of-mass energy of 5.44 TeV was carried out at the LHC.
The mass number of xenon being roughly halfway between that of a proton and that of a lead nucleus, upcoming data from the Xe+Xe run offer a unique possibility to test the predictive power of the hydrodynamic framework under simple, though substantial, variations of the geometry of the quark-gluon plasma.

%Scaling laws in initial conditions+hydro
%affects both the initial
%density profile right after the collision, and the hydrodynamic
%response to these initial conditions. 
The effects of varying the mass number, $A$, of the colliding nuclei can be evaluated using scaling laws, which play an important role in fluid dynamics. 
Both the number of observed particles~\cite{Eremin:2003qn,Bozek:2016kpf,Loizides:2016djv,Zakharov:2016bob} and the volume of the fluid are proportional to $A$, so that the density of particles per unit volume is essentially independent of $A$.
Varying $A$ amounts, in a first approximation, to scaling all space-time variables by $A^{1/3}$. 
Now, ideal hydrodynamics  
is scale invariant:  
The distribution of temperature and fluid velocity is strictly unchanged upon such a linear rescaling. 
This implies that, for instance, to a good
approximation transverse momentum spectra up to $p_t\sim 2$~GeV should be identical in Xe+Xe and Pb+Pb collisions. 
Scale invariance, however, is a rough approximation, which is broken by several effects: 
\begin{itemize}

\item 
The surface thickness of a nucleus, $a\sim 0.5$~fm, is independent of $A$, so that the boundary of smaller nuclei is relatively less sharp. 
This leads to a smaller eccentricity in the reaction plane for smaller systems~\cite{Alver:2006wh}, which in turn implies a smaller elliptic flow in mid-central collisions. 

\item
Short-range fluctuations~\cite{Takahashi:2009na} of the initial density profile. 
Their spatial extension is determined by the microscopic collision dynamics and is likely to depend little on $A$. 
Their effect is typically proportional to $A^{-1/2}$~\cite{Bhalerao:2011bp}. 
They explain why elliptic flow is sizable even in central 
collisions, and why it is larger for smaller $A$~\cite{Alver:2006wh}. 
These fluctuations are also responsible for triangular
flow~\cite{Alver:2010gr}. 

\item
The viscous corrections to ideal hydrodynamics,\footnote{By viscous
  corrections, we mean here {\it all\/} departures from local equilibrium: in the
  hydrodynamic modeling, they are due not only to the viscosity
  during the fluid phase, but also to the traditional
  freeze-out procedure~\cite{Cooper:1974mv}, which effectively takes
  into account the departure from equilibrium at the end of the 
  evolution~\cite{Borghini:2005kd}.} 
which involve gradients~\cite{Baier:2007ix}, and whose effects are, therefore, proportional to $A^{-1/3}$. 
Their main effect is to decrease the hydrodynamic response to the initial anisotropies~\cite{Romatschke:2007mq}. 
\item
The $^{208}$Pb nucleus is spherical while $^{129}$Xe has a moderate prolate deformation~\cite{Moller:2015fba}.
\end{itemize}
Precise comparisons between Pb+Pb and Xe+Xe data will provide an exceptional opportunity of verifying these scaling rules, which lie at the heart of the hydrodynamic modeling.

%Why we can make predictions without knowing initial conditions and viscosity
The main limitations of the hydrodynamic framework are the poor knowledge of the initial density profile, and of the unknown transport coefficients (viscosity) of the quark-gluon plasma~\cite{Luzum:2008cw,Shen:2015msa}.  
%Current efforts from Lattice QCD \cite{Mages:2015rea} may one day determine the viscosity from first principles, however, in the meantime the best efforts arise from hydrodynamics comparisons to experimental data. 
%JYO: this is not the point of this paper at all. 
Some models of the initial density can be ruled out by combining elliptic and triangular flow data~\cite{Retinskaya:2013gca}, in the sense that they do not yield good descriptions of experimental data even after tuning the viscosity.
Nevertheless, even if a particular model of initial conditions can be made compatible with data at the expense of adjusting the viscosity, it is not guaranteed that it provides an accurate representation of reality.\footnote{A model may, for instance, underestimate both the initial eccentricity and the density fluctuations. A good descsription of data could be achieved, then, through an overestimation of the hydrodynamic response to the initial anisotropies, i.e., by implementing a smaller viscosity.}

We can argue, though, that if a particular hydrodynamic calculation matches Pb+Pb data, it should correctly predict Xe+Xe data, even if it has the wrong initial conditions and viscosity. 
The uncertainty in initial conditions comes from the microscopic dynamics, not from the structure of the nuclei: 
If a model of initial conditions overestimates both eccentricity and fluctuations in Pb+Pb collisions, it is likely to also overestimate them by the same fraction in Xe+Xe collisions. 
It should be stressed that the error on initial conditions will not be exactly compensated by the error on the viscosity, because viscous damping is larger in smaller systems. 
However, this is a small change, because viscous
effects are proportional to $A^{-1/3}$, so that they increase only by 17\% from $^{208}$Pb to $^{129}$Xe. 
Therefore, even though we do not have control over all the features of the hydrodynamic modeling, we are in condition of making robust predictions for the system size dependence of typical hydrodynamic signatures. 
%By focusing on the size dependence, the uncertainties are significantly reduced --- not only theoretical uncertainties, but also experimental systematic uncertainties, which will partially cancel for pairs of measurements done at the two collision energies using the same detector and with the same analysis.

We focus on a few bulk, soft observables, which allow to test directly the hydrodynamic scaling: Mean transverse momentum, anisotropic flow, and flow fluctuations.
We do not study in detail the dependence of yields or flow on transverse momentum, pseudorapidity, or particle type.
Simulations are carried out for both Pb+Pb collisions at $\sqrt{s_{\rm NN}}=5.02$ TeV, and Xe+Xe collisions at $\sqrt{s_{\rm NN}}=5.44$ TeV.
In Sec.~\ref{s:initial}, we present the details of the initial condition model used in this paper, and the setup of our hydrodynamic code.
Results are presented in Sec.~\ref{s:results}.

\section{Hydrodynamic modeling}
\label{s:initial}

Since we do not study the rapidity dependence of observables, we assume that the longitudinal expansion of the medium is boost
invariant~\cite{Bjorken:1982qr}.
In each event we specify the initial density profile over the transverse plane, and then determine the transverse expansion 
numerically using a 2+1 dimensional hydrodynamic code. 

\subsection{Initial state}

The model of initial conditions that we shall use throughout this paper is the  {\trento} 
model with %reduced thickness parameter 
$p=0$ (we refer to \cite{Moreland:2014oya} for a detailed description of this model).
In this model, the total entropy deposited at a given point in the transverse plane after the collision is calculated as $\sqrt{T_AT_B}$, where $T_A$ and
$T_B$ are the thickness functions of the incoming nuclei at that point.\footnote{The Woods-Saxon parameterization used for shaping the $^{129}$Xe nuclei is taken from Ref.~\cite{Loizides:2014vua} for what concerns the surface thickness, $a$, and the nuclear radius, $R$. The shape parameters characterizing the deformation of the nucleus are instead taken from Ref.~\cite{Moller:2015fba}. Thus, our Xe nuclei present $A=129$, $R=5.42$ fm, $a=0.57$ fm, $\beta_2=0.162$, $\beta_4=-0.003$.}
In the remainder of this section we shall explain that this choice of initial state model is strongly motivated by experimental results on Pb+Pb collisions.

\subsubsection{Multiplicity and centrality}
\begin{figure}[t!]
\includegraphics[width=.9\linewidth]{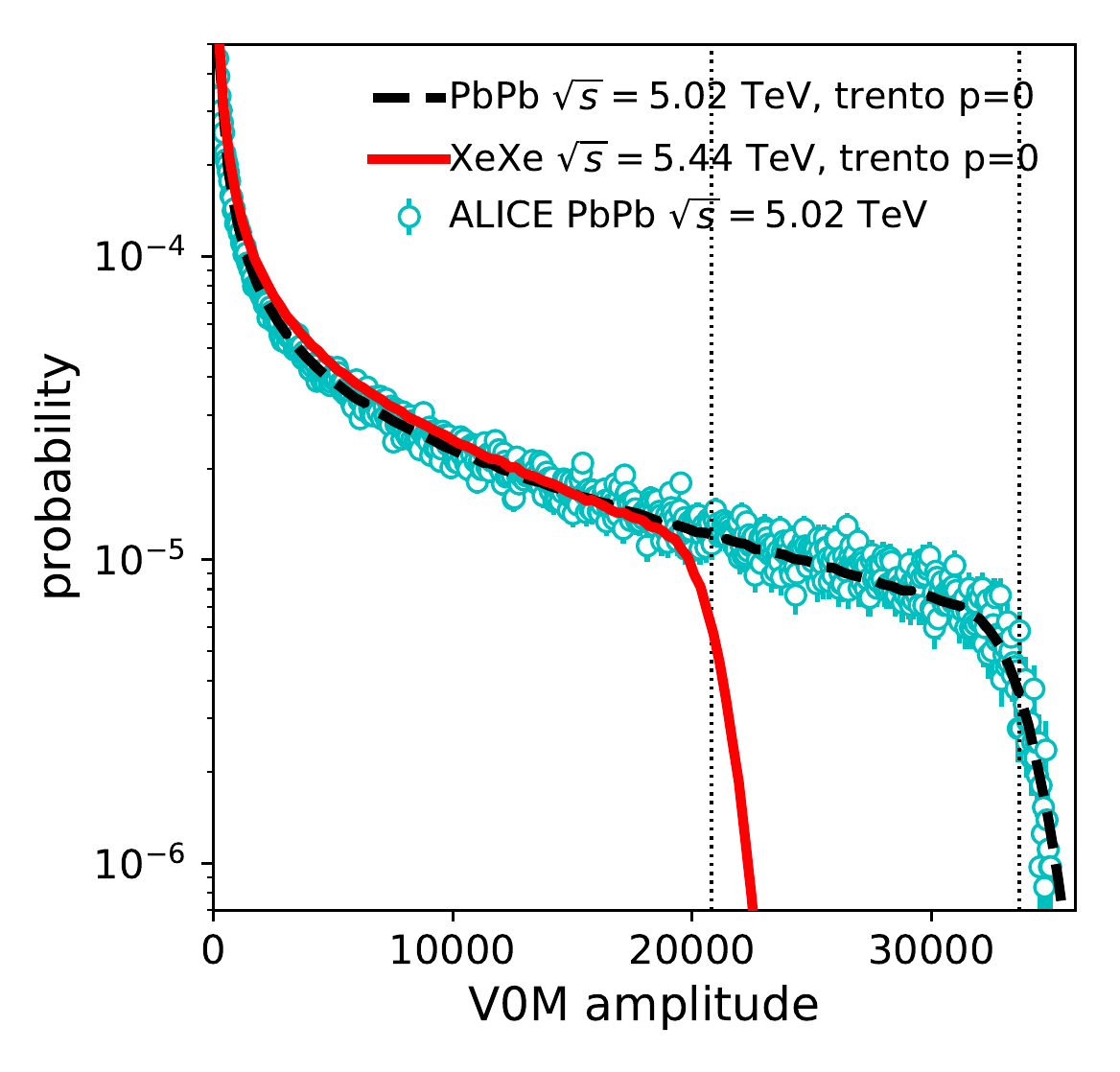}
\caption{\label{fig:centrality} Probability distribution of the
V0M amplitude, used by the ALICE Collaboration to sort events into centrality classes~\cite{ALICEnote}. Symbols: ALICE data for $5.02$~TeV Pb+Pb
  collisions. Dashed line: Rescaled {\trento} entropy in $5.02$~TeV Pb+Pb
 events. Full line: Prediction for $5.44$~TeV
  Xe+Xe collisions using the same model. The vertical dotted lines indicate the locations of the knees of the histograms (see text for more details).
} 
\end{figure}
In experiment, the centrality of a nucleus-nucleus collision is
defined according to the 
transverse energy~\cite{ATLAS:2011ah,Chatrchyan:2011pb} or 
multiplicity~\cite{Abelev:2013qoq} in a detector. 
While these are, strictly speaking, final-state observables, they
mostly reflect the initial entropy of the medium for the following reasons: 
The entropy increase due to viscosity during the hydrodynamic phase is
typically a small fraction, and the transverse energy per particle
depends little on centrality. 
Therefore, one can assume that the quantity used to determine the
centrality is proportional to the initial entropy.
Within this assumption, the \trento{} model with $p=0$ is very successful in reproducing distributions of multiplicity measured at the LHC in Pb+Pb collisions \cite{Moreland:2014oya}.
We show in Fig.~\ref{fig:centrality} the distribution of entropy obtained in this model, and we compare it to the distribution of the V0M amplitude used by ALICE to sort events into centrality classes, in Pb+Pb collisions at $\sqrt{s_{\rm NN}}=5.02$ TeV \cite{ALICEnote}.
The values of entropy provided by \trento{} are rescaled on the horizontal axis, so that the histogram of the model and that of ALICE data present the same \textit{knee}, which, following \cite{Das:2017ned}, is defined as the mean value of the V0M amplitude at zero impact parameter.\footnote{Alternatively, one can define the knee as the rightmost inflection point if one plots the probability of Fig.~\ref{fig:centrality} in a linear scale, instead of a logarithmic scale~\cite{ATLAS:2017zcm}. Both methods are essentially equivalent.}
Once the knees are matched, histograms are in perfect agreement. 
To be more quantitative, we check that the fraction of events on the right of the knee, i.e., the centrality of the knee, $c_{\rm knee}$, is correctly reproduced by the model.\footnote{The \trento{} calculation shown in Fig.~\ref{fig:centrality} implements fluctuation parameter $k=2.0$, which is found to provide the best agreement with the measured $c_{\rm knee}$. The \trento{} events that shall be used as initial condition for the hydrodynamic evolution, on the other hand, present $k=1.6$, as suggested in \cite{Bernhard:2016tnd}. This value gives a $c_{\rm knee}$ that is slightly too large, but the difference has a negligible effect on the observables presented in Sec.~\ref{s:results}.}
Indeed, we obtain $c_{\rm knee}=0.39\pm0.02 \%$ in \trento{}, while ALICE data present $c_{\rm knee}=0.38\pm0.04\%$.
We stress that this quantity is independent of the calibration of the measured multiplicity (horizontal axis).

A second nontrivial success of this model is that it captures correctly the dependence of multiplicity on the system size.
We can check this explicitly using the Relativistic Heavy-Ion Collider (RHIC) data on Cu+Cu and Au+Au collisions at $\sqrt{s_{\rm NN}}=200$ GeV.
The $p=0$ model predicts the multiplicity of central Au+Au collisions to be 3.5 times larger that that of a central Cu+Cu collisions.
This is a bit larger than scaling expected from the ratio of the mass numbers, $197/63\simeq 3.1$, and turns out to be in agreement with RHIC data, as the multiplicity of charged particles measured in Au+Au collisions is larger by a factor 3.8~\cite{Alver:2010ck}.

Motivated by these features, we show in Fig.~\ref{fig:centrality} the prediction of the $p=0$ model for the histogram of the V0M amplitude in Xe+Xe collisions at $~\sqrt{s_{\rm NN}}=5.44$ TeV.
We predict $c_{\rm knee}=0.51\pm0.04\%$ in Xe+Xe collisions.
Note that, the centrality of the knee being due to fluctuations, it follows to a good extent the expected $A^{-1/2}$ scaling, as $\sqrt{208/129}\approx 0.51/0.39$.

\begin{figure}
\includegraphics[width=.85\linewidth]{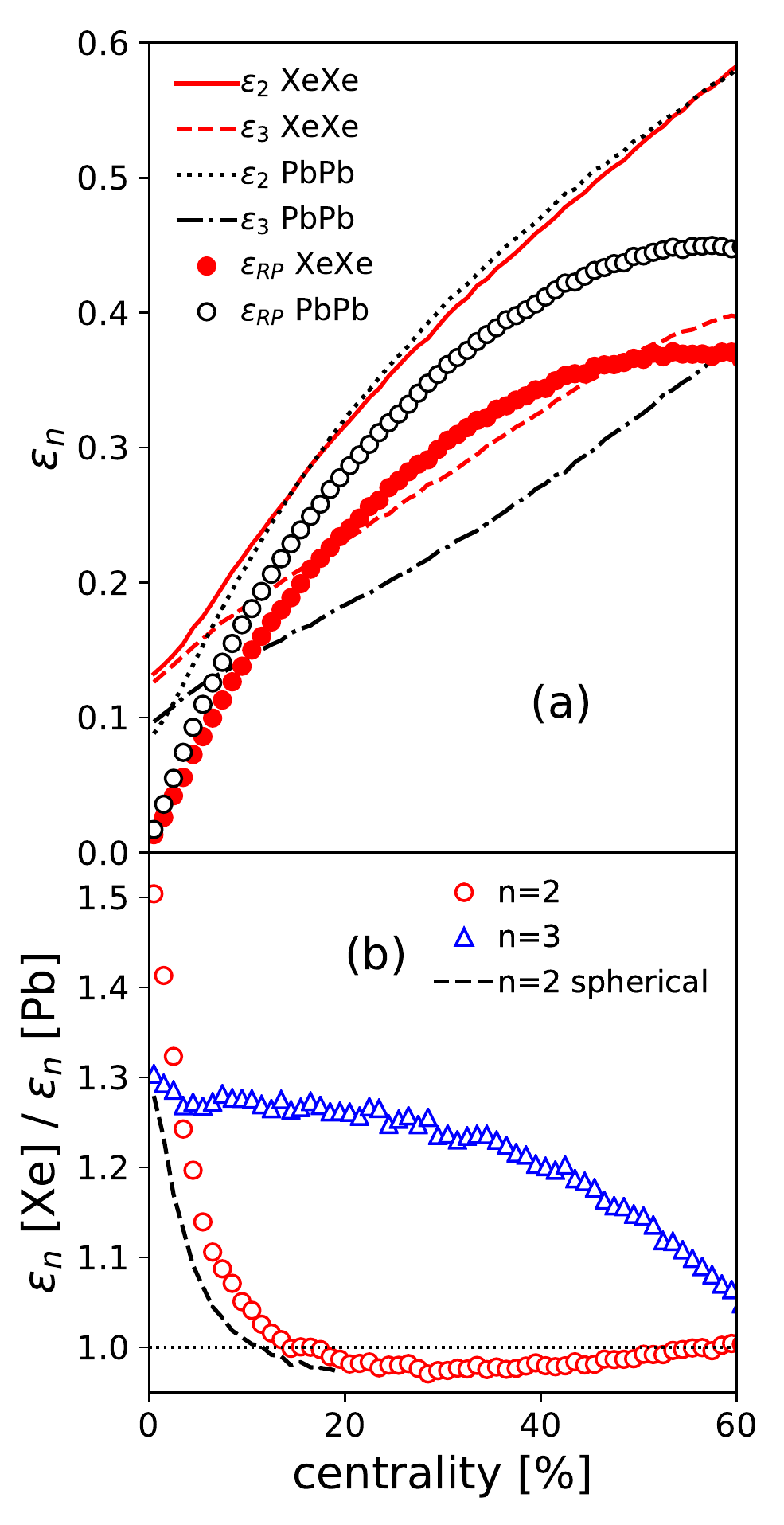}
\caption{\label{fig:ecc} (a) Root-mean-square values of the initial
  eccentricity, $\varepsilon_2$, and the initial triangularity,
  $\varepsilon_3$~\cite{Qiu:2011iv}, in the \trento{} model, as a
  function of centrality percentile, in Pb+Pb collisions at $\sqrt{s_{\rm NN}}=5.02$ TeV, and Xe+Xe collisions at $\sqrt{s_{\rm NN}}=5.44$ TeV. 
$\varepsilon_{RP}$ denotes the mean value of the eccentricity in the
  reaction plane~\cite{Alver:2006wh}. 
(b) Ratio of rms $\varepsilon_2$ and $\varepsilon_3$ between Xe+Xe and Pb+Pb collisions. The dashed line in this panel is the result for $\varepsilon_2$ obtained by switching off the deformation of the Xe nucleus (see text).}
\end{figure}

\subsubsection{Initial anisotropies}
The other crucial aspect of the initial state model is that
it provides the medium with a geometry and its anisotropies, which are known to leave peculiar phenomenological signatures. 
In particular, the most prominent momentum anisotropies of the final-state particle distribution, elliptic flow and triangular flow, arise mainly from the eccentricity, $\varepsilon_2$, and triangularity, $\varepsilon_3$, of the system \cite{Teaney:2010vd}, respectively.
The $p=0$ model is known to yield anisotropies which are compatible with experimental data, in the sense that it presents rms $\varepsilon_2$ and $\varepsilon_3$ which pass the test proposed in Ref.~\cite{Retinskaya:2013gca} across the full centrality range.
Further, relative fluctuations of $\varepsilon_2$ in this model turn out to be in good agreement with data on multi-particle cumulants of elliptic flow in central collisions~\cite{Giacalone:2017uqx}.
In Fig.~\ref{fig:ecc}--(a), we present the rms eccentricites and triangularities predicted by the \trento{} model as function of centrality percentile, in both Pb+Pb and Xe+Xe collisions. 

The triangular anisotropy of the medium, $\varepsilon_3$, is not expected to be dynamically generated by the fluctuations of orientations of the deformed $^{129}$Xe nuclei, and it is solely due to density fluctuations in both Xe+Xe and Pb+Pb.
%\footnote{A convenient measure of the system size, which can be
%  used to compare different colliding systems, is the charged
%  multiplicity~\cite{Basar:2013hea}.} 
Therefore, it is larger if the system is smaller, and it increases with the centrality percentile for a given
system, and it is larger in Xe+Xe collisions than 
in Pb+Pb collisions at the same centrality percentile. 
Figure~\ref{fig:ecc}--(b) shows that the ratio between the initial triangularities of Xe+Xe and Pb+Pb collisions is close to 1.27 for most centralities, in agreement with the expected $A^{-1/2}$ scaling. 

The eccentricity, $\varepsilon_2$, on the other hand, gets contribution from both fluctuations
and the almond shape of the system due to finite impact parameter, which is quantified by the mean eccentricity in the reaction plane, dubbed 
$\varepsilon_{\rm RP}$ in the figure.
Fluctuations and $\varepsilon_{\rm RP}$ add in quadrature~\cite{Voloshin:2007pc}: 
$\varepsilon_2=\sqrt{\varepsilon_{\rm RP}^2+\sigma^2}$, where $\sigma^2$
denotes the variance of $\varepsilon_2$ fluctuations. 
The reaction plane eccentricity, $\varepsilon_{\rm RP}$, is larger in
Pb+Pb than in Xe+Xe at all centralities, which is explained by the
sharper nuclear surface of the Pb+Pb system.
In central collisions, however, $\varepsilon_{\rm RP}$ vanishes and 
$\varepsilon_2$ is solely due to fluctuations: It is of the same order
as $\varepsilon_3$ in both systems.
Hence, $\varepsilon_2$ is larger in Xe+Xe than Pb+Pb for central
collisions, but smaller for mid-central collisions [Fig.~\ref{fig:ecc}--(b)].  
Note that in very central collisions the ratio of the $\varepsilon_2$ coefficients presents a significant deviation from the value 1.27 expected from $A^{-1/2}$ scaling.
This is due to the prolate shape of the $^{129}$Xe nuclei, in particular, to a nonzero value of the parameter $\beta_2$.
The random orientation of the colliding nuclei provides a dynamical source of eccentricity fluctuations, which dominates over density fluctuations in central collisions, as known, for instance, from experimental investigations of $^{238}$U+$^{238}$U collisions \cite{Adamczyk:2015obl}.
For completeness, then, the dashed line in Fig.~\ref{fig:ecc}--(b) indicates the ratio of eccentricities that one would obtain if $^{129}$Xe nuclei were spherical, i.e., by setting $\beta_2=\beta_4=0$ in the Woods-Saxon parametrization.
As expected, this ratio in central collisions naturally follows the $A^{-1/2}$ scaling. 
The effect of deformation is therefore maximum at 0\% centrality, where it increases by  nearly 20\% the value of $\varepsilon_2$, and disappears around 20\% centrality.

\subsection{Hydrodynamic evolution}
We sort a sample of few million \trento{} simulations ($p=0$, $k=1.6$, $\sigma=0.51$) of both Pb+Pb collisions at $\sqrt{s_{\rm NN}}=5.02$~TeV, and Xe+Xe collisions at $\sqrt{s_{\rm NN}}=5.44$~TeV into centrality classes, through bins of 5\% width, from 0\% to 60\% centrality.
In each bin we evolve hydrodynamically approximately 2200 events, for both systems.
Each initial density profile is evolved by means of the viscous 
relativistic hydrodynamical code {\footnotesize V-USPHYDRO}
\cite{Noronha-Hostler:2013gga,Noronha-Hostler:2014dqa,Noronha-Hostler:2015coa}.

The equation of state (PDG16+/2+1[WB]) we use is that recently calculated on the lattice with three quark flavors ($u$, $d$, $s$) and physical quark masses~\cite{Borsanyi:2013bia}.
%\cite{Borsanyi:2016ksw}
We start the hydrodynamic evolution at a time $\tau_0=0.6$~fm/c after
the time of collision~\cite{Kolb:2000fha}. 
We neglect the transverse expansion before 
$\tau_0$~\cite{Vredevoogd:2008id,vanderSchee:2013pia,Liu:2015nwa,Keegan:2016cpi}. 
Doing so, we underestimate the transverse flow, which we partially compensate with the implementation of a low
shear viscosity over entropy ratio, $\eta/s=0.047$~\cite{Alba:2017hhe}. 
The bulk viscosity~\cite{NoronhaHostler:2008ju,Monnai:2009ad,Bozek:2009dw,Noronha-Hostler:2013gga,Noronha-Hostler:2014dqa} is set to zero in our calculation. 
These parameters are the same used in Ref.~ \cite{Alba:2017hhe}, and were chosen so as to reproduce experimental data on Pb+Pb collisions at $\sqrt{s_{\rm NN}}=5.02$ TeV.
We run hydrodynamics until the temperature drops below $150$~MeV, at
which point the fluid transforms into hadrons~\cite{Teaney:2003kp}.  
The equation of state of the fluid matches to a hadron resonance gas model with the most up-to-date particles from the Particle Data Group that includes all *-**** resonances \cite{Patrignani:2016xqp}, which are shown to be needed from first principles \cite{Alba:2017mqu}.
All these hadronic resonances can be formed during the freeze-out process.
We neglect rescatterings in the hadronic phase~\cite{Song:2011hk}, but implement strong decays of all hadronic resonances into stable hadrons using an adapted version of the decay code of \cite{Kolb:2002ve}. 
%add references to explain which resonances are included and how
%decays are done.
\begin{figure}[t!]
\includegraphics[width=\linewidth]{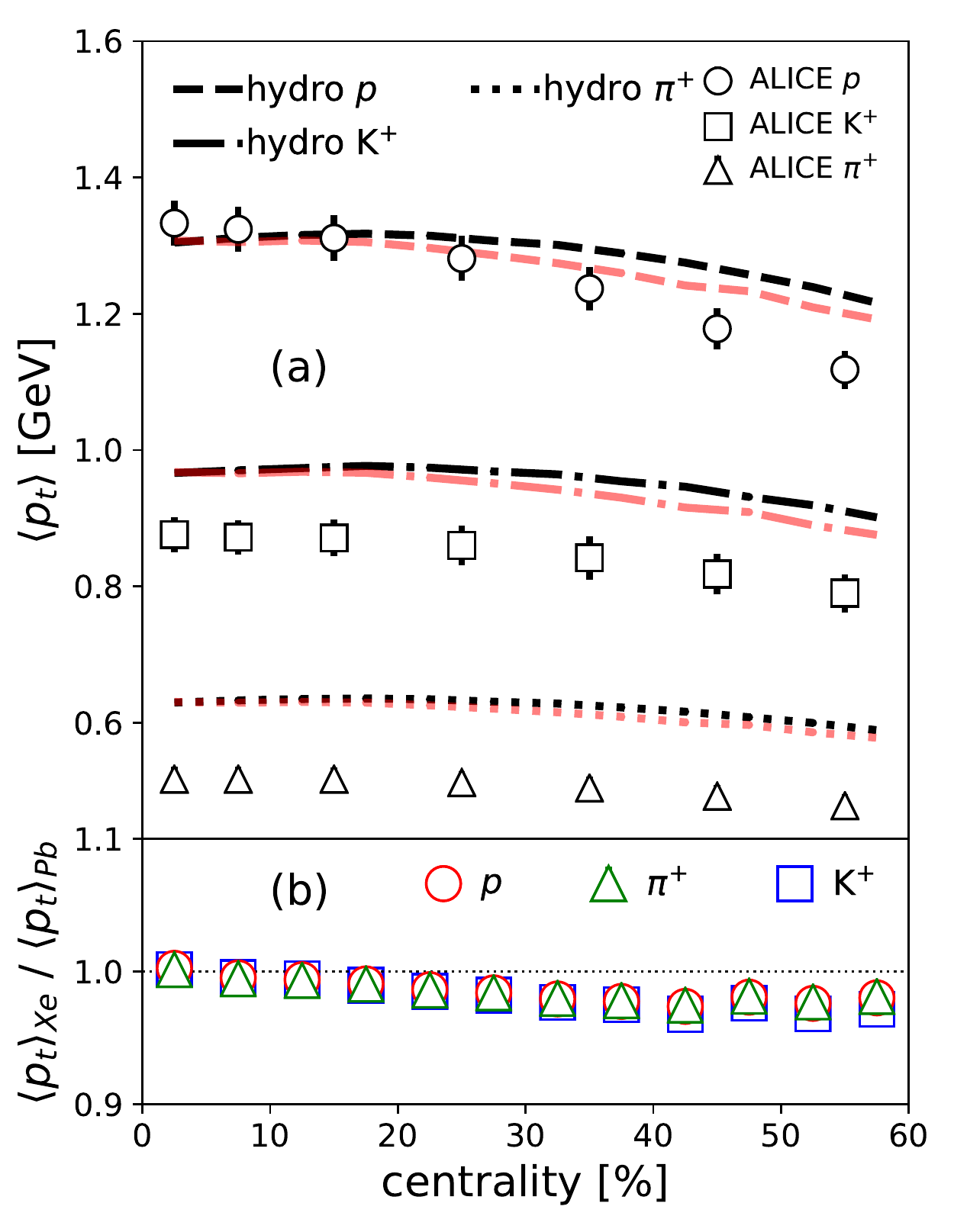}
\caption{\label{fig:pt} 
(color online) (a) Mean transverse momentum, $\langle p_T\rangle$, of identified particles as a function of
  centrality percentile. 
Symbols: ALICE results for Pb+Pb collisions at $\sqrt{s_{\rm
    NN}}=2.76$~TeV~\cite{Abelev:2013vea}. 
Black lines: Our hydrodynamic results for Pb+Pb collisions at
$\sqrt{s_{\rm NN}}=5.02$~TeV. 
Grey lines: Our hydrodynamic results for Xe+Xe collisions at
$\sqrt{s_{\rm NN}}=5.44$~TeV. 
Panel (b) displays the ratio between Xe+Xe and Pb+Pb
collisions. 
}
\end{figure}

\section{Results}
\label{s:results}

\subsection{Transverse momentum spectra}

Figure~\ref{fig:pt}--(a) displays the mean transverse momentum of pions,
kaons, and protons in our hydrodynamic calculation as a function of
the centrality percentile.  
We present our results along with ALICE data at a lower energy~\cite{Abelev:2013vea}, since
identified particle spectra at $\sqrt{s_{\rm NN}}=5.02$ TeV have not been published yet.

The mean transverse momentum is remarkably independent of centrality,
both in hydrodynamics and in experiment.
This flatness of data is a robust prediction of the hydrodynamic framework, and the fact that it is observed experimentally strongly supports the validity of the hydrodynamic approach. 
We find that the absolute value of $\langle p_{\rm T}\rangle$ is larger in our
calculation than in data, for pions and kaons. 
The discrepancy is larger
than that expected simply from the different colliding 
energy, which should yield a 3\% variation~\cite{Noronha-Hostler:2015uye}.
Agreement could be improved by adding a bulk 
viscosity~\cite{Ryu:2015vwa}.

However, our goal in this paper is to predict how 
$\langle p_T\rangle$ {\it evolves\/} between Pb+Pb and Xe+Xe collisions. 
The ratios of mean transverse momenta for different particle types are plotted in Fig.~\ref{fig:pt}--(b). 
The ratios are very close to unity, as expected from the scale invariance of
fluid dynamics, and the fact that the effective 
density~\cite{Monnai:2017cbv} is essentially identical in both
systems. 
Note that the increase of the energy per nucleon pair from 5.02 to
5.44~TeV between Pb+Pb and Xe+Xe collisions only has a small effect. 
It entails an increase of the multiplicity by $2.5\%$~\cite{Adam:2015ptt}
which, in turn, implies an increase of the mean transverse momentum by
$0.5\%$~\cite{Noronha-Hostler:2015uye}. 

\subsection{Anisotropies}

\begin{figure}
\includegraphics[width=\linewidth]{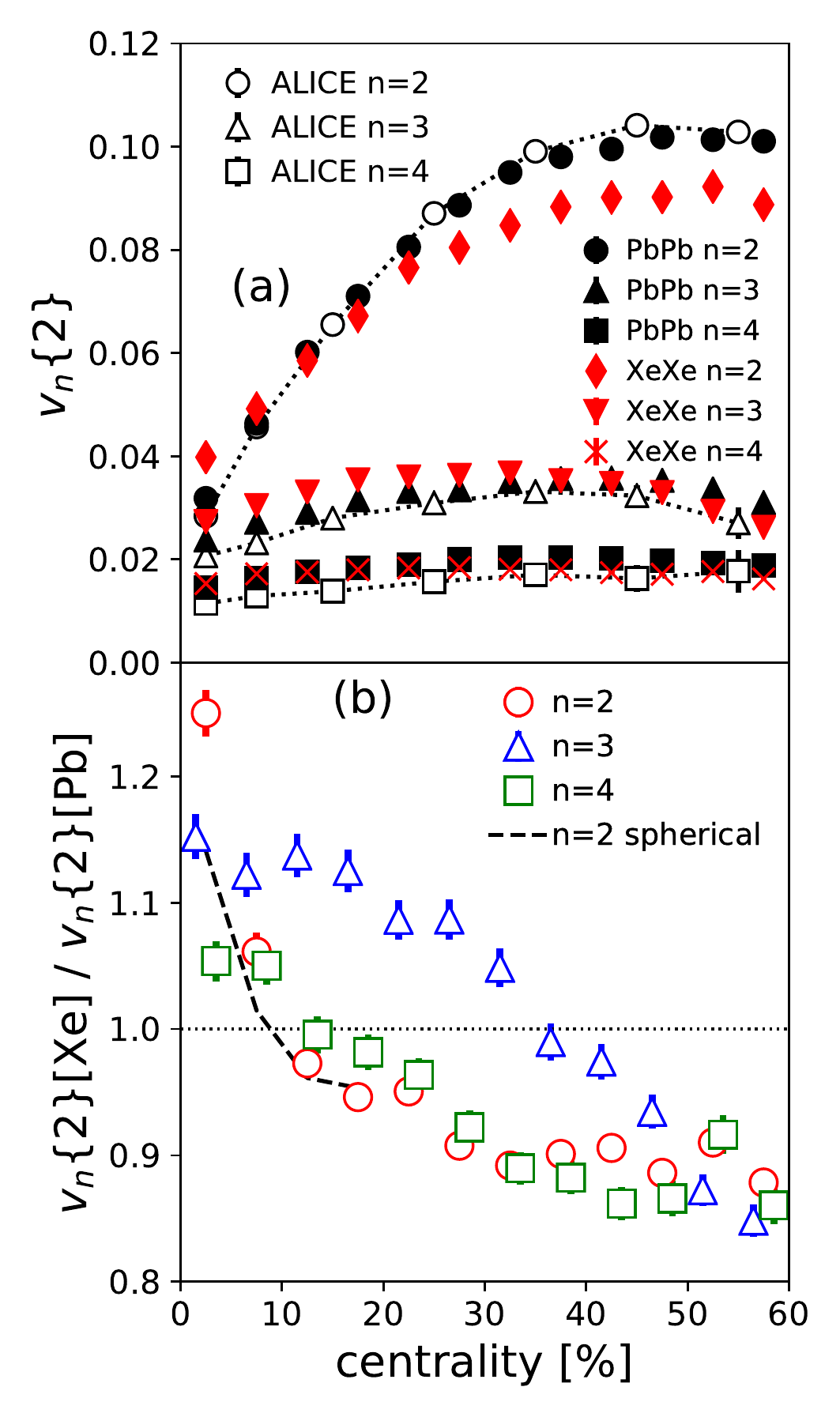}
\caption{\label{fig:vn} (a) rms values of $v_2$, $v_3$ and $v_4$ as a
  function of collision centrality. 
Open symbols: ALICE data for Pb-Pb collisions at $\sqrt{s_{\rm
    NN}}=5.02$~TeV~\cite{Adam:2016izf}. 
Full symbols: our hydrodynamic calculations for Pb+Pb and Xe+Xe
collisions. The kinematic cuts are $0.2<p_T<5$~GeV/c and $|\eta|<0.8$. 
%JYO: abstract of alice paper  https://arxiv.org/pdf/1602.01119.pdf
%The measurements are performed in the central pseudorapidity region |η| < 0.8 and for the transverse momentum range 0.2 < pT < 5 GeV/c.
(b) Hydrodynamic predictions for the ratios of the rms anisotropies between Xe+Xe and Pb+Pb collisions. 
The dashed in this panel is the result for $v_2$ obtained by switching off the deformation of the Xe nucleus. 
 }
\end{figure}

The anisotropies of the azimuthal distribution
of charged particles are quantified by the Fourier
coefficients $v_n$~\cite{Luzum:2011mm}, and represent a distinctive signature of collective
behavior. 
Figure~\ref{fig:vn}--(a) displays the rms value of $v_n$,
denoted by $v_n\{2\}$~\cite{Borghini:2001vi}, for $n=2,3,4$, as a
function of centrality percentile.  
Our calculation for Pb+Pb collisions is in good agreement with ALICE
data. 
It slightly overestimates $v_4$ and, to a lesser extent, $v_3$,
which could be improved by increasing the shear viscosity, as viscous
damping becomes more important for higher 
harmonics~\cite{Alver:2010dn,Luzum:2012wu}. 
Figure~\ref{fig:vn}--(b) displays our prediction for the ratios of
$v_n\{2\}$ between Xe+Xe and Pb+Pb collisions. 
We expect these ratios  to be independent of kinematic cuts in $p_t$ and $\eta$, and thus our prediction to be independent of the experimental setup.

Now, $v_2$ and $v_3$ are to a good approximation proportional to
$\varepsilon_2$ and $\varepsilon_3$~\cite{Niemi:2012aj}, i.e. $v_n=\kappa_n\varepsilon_n$. 
In ideal hydrodynamics, the response coefficient $\kappa_n$ is independent of the system 
size and shape. 
Viscous corrections
decrease $\kappa_n$ by an amount proportional to $1/R$, where $R$ is
the transverse size~\cite{Gombeaud:2007ub}. Hence, for a given system $\kappa_n$
decreases as a function of centrality percentile,
whereas at the same centrality it is smaller in a smaller system. 

For elliptic flow, $n=2$, the ratio in Fig.~\ref{fig:vn}--(b)
follows qualitatively the same 
trend as the ratio of values of $\varepsilon_2$ in Fig.~\ref{fig:ecc}--(b): 
It is larger than 1 in the most central bins, which is due to the
larger eccentricity fluctuations. It drops below 1 above 10\%
centrality, because of the smaller eccentricity in the reaction plane.
We note that in central collisions the ratio of the $v_2$ coefficients is much smaller than the corresponding ratio for the spatial eccentricities of Fig.~\ref{fig:ecc}--(b).
This is an effect of viscous damping, which affects more strongly the smaller system, Xe+Xe.
The situation is similar in peripheral collisions: The ratio of the initial eccentricities is close to unity above 50\% centrality, while the $v_2$ ratio in hydro remains around 0.9, indicating larger viscous damping in the smaller system.
Again, we show as a dashed line the ratio that one would obtain after hydrodynamic evolution of collisions of spherical $^{129}$Xe nuclei: It is of the same order as the ratio of the $v_3$ coefficients (triangles) in very central collisions, and matches the $v_2$ ratio with deformed nuclei (circles) around 15\% centrality.

Moving on to triangular flow, $n=3$, the ratio in Fig.~\ref{fig:vn}--(b) is
larger than 1 up to 40\% centrality, in agreement with the ratio of the corresponding triangularities [Fig.~\ref{fig:ecc}--(b)].
Above 40\% centrality, however, viscous damping causes the $v_3$ ratio to drop below 1, even though the ratio of $\varepsilon_3$ does not. 

Finally, the situation for $v_4$ is intermediate, as $v_4$ is driven by both fluctuations in the most central bins (ratio larger than unity),
and nonlinear coupling to $v_2$ in mid-central and peripheral
collisions~\cite{Gardim:2011xv} (ratio below unity). 
The latter feature also explains why $v_4$ is not affected by nuclear deformation effects.

\begin{figure}[t!]
\includegraphics[width=.82\linewidth]{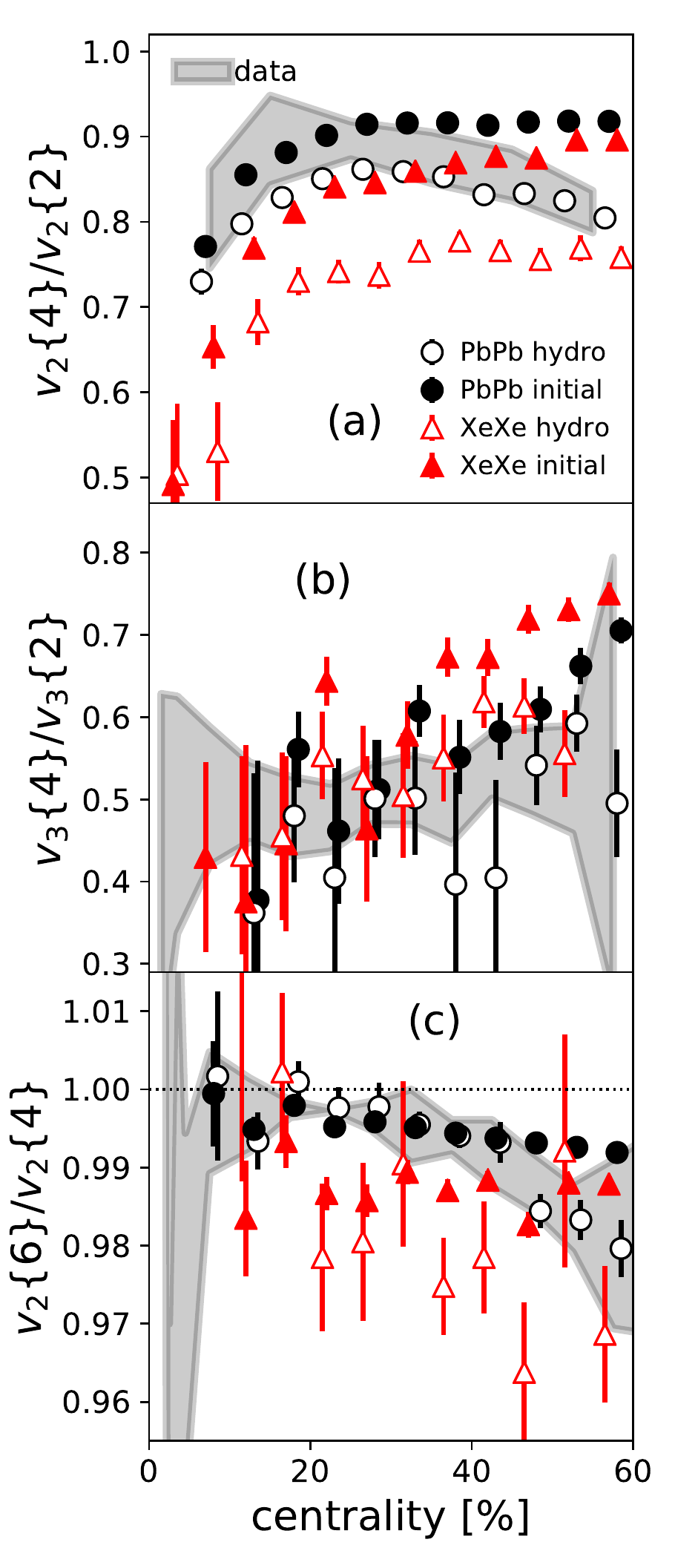}
\caption{\label{fig:cumulants}Ratios of cumulants of anisotropic flow
  (open symbols) and corresponding ratios for initial eccentricity harmonics
  (full symbols), in Pb+Pb and Xe+Xe collisions from our hydrodynamic
  calculation, as a function of centrality percentile. 
From top to bottom: (a) $v_2\{4\}/v_2\{2\}$ and
$\varepsilon_2\{4\}/\varepsilon_2\{2\}$. 
Data are from 5.02~TeV Pb+Pb
collisions collected by the ALICE Collaboration~\cite{Adam:2016izf}.
(b) $v_3\{4\}/v_3\{2\}$ and
$\varepsilon_3\{4\}/\varepsilon_3\{2\}$. Data are from 2.76~TeV Pb+Pb
collisions collected by the ATLAS Collaboration~\cite{Aad:2014vba}.
(c) $v_2\{6\}/v_2\{4\}$ and
$\varepsilon_2\{6\}/\varepsilon_2\{4\}$. 
Data are from 2.76~TeV Pb+Pb
collisions collected by the ATLAS Collaboration~\cite{Aad:2014vba}.
}
\end{figure}

\subsection{Flow fluctuations}

Cumulants of the probability distribution of anisotropic
flow~\cite{Borghini:2001vi} are
measured to a great accuracy~\cite{Abelev:2014mda,Aad:2014vba,Sirunyan:2017fts},
and give detailed insight into the statistical properties of 
$v_n$~\cite{Giacalone:2016eyu}. The first three cumulants of $v_n$ fluctuations, which we expect to be experimentally accessible after the Xe+Xe run at the LHC, are defined by 
\begin{eqnarray}
\nonumber v_n\{2\}^2 &=& \langle v_n^2 \rangle, \cr
\nonumber v_n\{4\}^4 &=& 2 \langle v_n^2\rangle^2 - \langle v_n^4\rangle, \\
\nonumber v_n\{6\}^6 &=& \frac{1}{4} \left( \langle v_n^6 \rangle - 9 \langle v_n^2\rangle \langle v_n^4 \rangle + 12 \langle v_n^2\rangle^3\right),
\label{eq:cumul} 
\end{eqnarray}
where angular brackets denote an average over events in a centrality bin. 

By taking ratios of cumulants of different orders in the same harmonic
$n$, one obtains direct information on the relative fluctuations of 
$v_n$~\cite{Giacalone:2017uqx}. 
The ratio $v_n\{4\}/v_n\{2\}$ is the simplest measure of relative
fluctuations: it is equal to 1 if $v_n$ is the same for all events, and smaller
than 1 otherwise. 
Figure~\ref{fig:cumulants}--(a) displays our hydrodynamic results for this ratio,
in the case $n=2$, as a function of the collision centrality. 
Results are presented along with ALICE data on Pb+Pb collisions at $~\sqrt{s_{\rm NN}}=5.02$ TeV \cite{Adam:2016izf}.
This ratio presents a distinctive feature: It first increases with centrality, and then changes trend.
This is correctly predicted by hydrodynamics, and our results are in fair agreement with data.
%$v_2$ 
%fluctuations are smallest for mid-central collisions, where elliptic
%flow is mostly due to the average almond-shaped geometry of the 
%overlap area between the two nuclei~\cite{Ollitrault:1992bk}. 
The remarkable point to note is that the ratio of cumulants of initial eccentricities, 
$\varepsilon_2\{4\}/\varepsilon_2\{2\}$, is somewhat larger than  
$v_2\{4\}/v_2\{2\}$, as a result of a nonlinear hydrodynamic 
response~\cite{Noronha-Hostler:2015dbi}. 
In particular, 
$\varepsilon_2\{4\}/\varepsilon_2\{2\}$ increases, and then saturates
as a function of centrality percentile. This is a
general feature of initial condition models~\cite{Giacalone:2017uqx}. 
Once the nonlinear hydrodynamic response is taken into account, however, one naturally recovers the non-monotonic behavior of $v_2\{4\}/v_2\{2\}$.
This shows that the success of the fluid-dynamical description of
elliptic flow extends beyond the linear response to the initial
eccentricity. 
Our prediction for Xe+Xe collisions is also shown in Fig.~\ref{fig:cumulants}--(a). 
At a given centrality, elliptic flow fluctuations are larger in Xe+Xe than in Pb+Pb, resulting in a smaller $v_2\{4\}/v_2\{2\}$ than for Pb+Pb collisions.
Note that the difference between $\varepsilon_2\{4\}/\varepsilon_2\{2\}$ and  $v_2\{4\}/v_2\{2\}$, due to nonlinear hydrodynamic response, is larger for Xe+Xe than for Pb+Pb. 

Figure~\ref{fig:cumulants}--(b) presents the ratio
$v_3\{4\}/v_3\{2\}$, which is significantly smaller than
$v_2\{4\}/v_2\{2\}$, as triangular flow is solely due to 
fluctuations~\cite{Alver:2010gr}. 
%In fact, $v_3\{4\}$ vanishes if the probability distribution of $v_3$
%is a 2-dimensional Gaussian, so that the ratio $v_3\{4\}/v_3\{2\}$
%directly probes the non-Gaussianity of $v_3$
The ratio $v_3\{4\}/v_3\{2\}$
directly probes the non-Gaussianity of $v_3$ fluctuations~\cite{Giacalone:2017uqx,Abbasi:2017ajp}, and  
non-Gaussian fluctuations of the initial triangularity, $\varepsilon_3$,
are expected as a consequence of finite-size corrections to the
central limit theorem~\cite{Bhalerao:2011bp,Yan:2013laa,Gronqvist:2016hym}. 
On this basis, one expects $\varepsilon_3\{4\}/\varepsilon_3\{2\}$ to
scale with the number of participant nucleons, $N$, like $N^{-1/4}$. 
The results on this ratio provided by our initial condition models, displayed as full symbols in 
Fig.~\ref{fig:cumulants}--(b), follow the expected behavior:
$\varepsilon_3\{4\}/\varepsilon_3\{2\}$ increases as a function of
centrality percentile, and it is larger for Xe+Xe than for Pb+Pb. 
The
ratio between Xe+Xe and Pb+Pb is close to the value 1.13 expected on
the basis of $A^{-1/4}$ scaling.
Note that the value of $v_3\{4\}/v_3\{2\}$ at the end of the hydrodynamic
calculation is smaller than
$\varepsilon_3\{4\}/\varepsilon_3\{2\}$, in particular
above 30\% centrality. This results in a flatter centrality dependence
for $v_3\{4\}/v_3\{2\}$ than for
$\varepsilon_3\{4\}/\varepsilon_3\{2\}$. 
These hydrodynamic results are consistent with experimental 
data~\cite{ALICE:2011ab,Chatrchyan:2013kba,Aad:2014vba}
which so far do not show any clear centrality dependence of
$v_3\{4\}/v_3\{2\}$ for Pb+Pb collisions. 
We do not seize any clear difference between the values of $v_3\{4\}/v_3\{2\}$ in Xe+Xe and in Pb+Pb collisions.

Finally, Figure~\ref{fig:cumulants}--(c) displays the ratio
$v_2\{6\}/v_2\{4\}$. This ratio is equal to 1 if the fluctuations of
$v_2$ are Gaussian~\cite{Voloshin:2007pc}. Hydrodynamics predicts
its value to be slightly smaller than 1. This originates from the fact that the
eccentricity in the reaction plane is bounded by unity, which skews
the distribution of $\varepsilon_2$~\cite{Giacalone:2016eyu}. 
The ratio $v_2\{6\}/v_2\{4\}$ is close to the corresponding ratios for
initial eccentricities $\varepsilon_2\{6\}/\varepsilon_2\{4\}$ for the
most central bins and then becomes gradually smaller. 
Interestingly, experimental data for Pb+Pb
collisions~~\cite{Abelev:2014mda,Aad:2014vba,Sirunyan:2017fts} are in
perfect agreement with the full hydrodynamic 
calculation, and not with the ratio $\varepsilon_2\{6\}/\varepsilon_2\{4\}$ from the initial state. This 
underlines once more that the success of hydrodynamics goes beyond linear
response to $\varepsilon_2$. 
We predict a smaller value of $v_2\{6\}/v_2\{4\}$ in Xe+Xe collisions
than in Pb+Pb collisions for all centralities.

Before concluding, we stress that we have explicitly checked the impact of the nuclear deformation on the presented ratios of cumulants.
We find that the effect of switching off the deformation of the $^{129}$Xe nuclei is smaller than the statistical error bars shown in Fig.~\ref{fig:cumulants}, and, therefore, negligible in our calculation.

\section{Conclusions}
We have presented predictions for upcoming data on Xe+Xe collisions 
at $\sqrt{s_{\rm NN}}=5.44$~TeV at the LHC.
Detailed comparisons with results from Pb+Pb collisions at $\sqrt{s_{\rm NN}}=5.02$~TeV
provide a unique opportunity to 
directly test fundamental scaling rules obeyed by hydrodynamic models. 
The mean transverse momentum is the same in Xe+Xe as
in Pb+Pb collision, the difference being at most 2\% in
mid-central collisions. 
Elliptic flow is larger by 25\% in Xe+Xe (where 15\% are due to the change in system size and 10\% to the deformation of the $^{129}$Xe nucleus) than in Pb+Pb in
the 0-5\% centrality window, but smaller by 10\% above 30\%
centrality. 
Triangular flow is larger than in Pb+Pb collisions up to 30\%
centrality, and smaller above 40\% centrality. 
The maximum value of the ratio $v_2\{4\}/v_2\{2\}$
is around 0.8 in Xe+Xe collisions, while it reaches 0.9 in Pb+Pb
collisions. 
The relative difference between $v_2\{4\}$ and $v_2\{6\}$ is also
significantly larger, typically by a factor 2, in
Xe+Xe than in Pb+Pb collisions. 
The latter predictions on cumulant ratios probe the ability of 
hydrodynamics to model anisotropic flow in the 
non-linear regime, beyond linear response to initial anisotropies.

\section{Acknowledgments}
We thank the experimental physicists working at the CERN LHC for the useful input.
We thank Alberica Toia for providing us with the ALICE data used in
Fig.~\ref{fig:centrality}. 
J.N.H. acknowledges the Office of Advanced Research Computing (OARC) at Rutgers, The State University of New Jersey for providing access to the Amarel cluster and associated research computing resources that have contributed to the results reported here. J.N.H also acknowledges the use of the Maxwell Cluster and the advanced support from the Center of Advanced Computing and Data Systems at the University of Houston. 
This work is funded under the USP-COFECUB project Uc Ph 160-16
(2015/13) and under the FAPESP-CNRS project 2015/50438-8.  
ML acknowledges support from FAPESP projects 2016/24029-6  and 2017/05685-2, and project INCT-FNA Proc.~No.~464898/2014-5

\end{document}